\documentclass[ams, prb, twocolumn, amssymb,floatfix]{revtex4-1}
\usepackage{amsmath}
\usepackage{tabularx}
\usepackage{bm}
\usepackage{euscript}
\usepackage{graphicx}
\usepackage{color}
\usepackage{amsfonts}
\usepackage{exscale}
\usepackage{amsbsy}
\usepackage{subfigure}
\usepackage{textcomp}
\usepackage{comment}

\usepackage{hyperref}

\newcommand{\beq}{\begin{equation}}
\newcommand{\eeq}{\end{equation}}
\newcommand{\be}{\begin{eqnarray}}
\newcommand{\ee}{\end{eqnarray}}

\begin{document}

\title{Composite fermions and the field-tuned superconductor-insulator transition}
\author{Michael Mulligan$^1$ and S. Raghu$^{1,2}$}
\affiliation{$^{1}$Stanford Institute for Theoretical Physics, Stanford University, Stanford, California 94305, USA}
\affiliation{$^2$SLAC National Accelerator Laboratory, 2575 Sand Hill Road, Menlo Park, CA 94025, USA}
\date{\today}

\begin{abstract}
In several two-dimensional films that exhibit a magnetic field-tuned superconductor to insulator transition (SIT), stable metallic phases have been observed.  
Building on the `dirty boson' description of the SIT, we suggest that the metallic region is analogous to the composite Fermi liquid observed about half-filled Landau levels of the two-dimensional electron gas.  
The composite fermions here are mobile vortices attached to one flux quantum of an emergent gauge field.  
The composite vortex liquid is a 2D non-Fermi liquid metal, which we argue is stable to weak quenched disorder.  
We describe several experimental consequences of the emergent composite vortex liquid.   
\end{abstract}

\maketitle

\section{Introduction}
How is superconductivity destroyed by a perpendicular magnetic field at $T=0$ in a disordered thin film \cite{Goldman2010, Hebard1990,Paalanen1992, Yazdani1995, Gantmakher1998, MasonKapitulnik1999, Mason2001, Bielejec2002, Sambandamurthy2004, Steiner2005, Qin2006, Steiner2008, Breznay2015}?  
There are two approaches to addressing this question.  
The first is a fermionic description where the destruction of superconductivity is identified as the loss of superconducting pairing amplitude \cite{Feigelman2001}.  
The second, so called `dirty boson' approach, assumes the magnetic field induces a superconductor to insulator transition (SIT), where Cooper-pair {\it localization} occurs, while the amplitude remains finite across the transition \cite{Fisher1990,Fisher1990a}.  

In the presence of {\it both} disorder and a strong magnetic field ($H> H_{c1}$), superconductors and insulators are sharply defined only at $T=0$ in 2D \cite{FFH1991}.
Thus, the field-tuned SIT is a rather unconventional quantum critical point.
In contrast, the zero-field disorder-tuned SIT \cite{Goldman2010, Markovic1998} has a line of finite-temperature superconducting transitions that terminate at the zero-field disorder-tuned SIT \cite{Abrikosov1958,Anderson1959}.
This, along with the preservation of time-reversal symmetry, implies that the zero-field disorder-tuned SIT {\it must} be in a distinct universality class. 
Our focus here will be on field-tuned transitions.  

In strongly disordered films, where the normal state resistance is well in excess of the Cooper-pair quantum of resistance, $R_Q = h/4e^2 \simeq 6.45k \Omega/ \square$, many predictions of the dirty boson theory have been confirmed \cite{Hebard1990, Paalanen1992, Steiner2005, Crane2007, Steiner2008, Breznay2015}.  
However, in somewhat cleaner films, a direct transition from superconducting to insulating behavior is lost.  
Instead, an intervening metallic phase with substantial superconducting fluctuations \cite{LiuPanWenKimSamArmitage} has been observed \cite{Yazdani1995, MasonKapitulnik1999, Mason2001, Steiner2005, Qin2006, Steiner2008}: the resistance approaches a constant, field-dependent value $R(T\rightarrow 0,H) \ll R_Q$ as $T \rightarrow 0$.   
To the extent that such metallic behavior observed at finite temperature uncovers the properties of a true zero temperature metal, it lies outside the scope of dirty bosons. 
\begin{figure}[h!]
  \centering
\includegraphics[width=.9\linewidth]{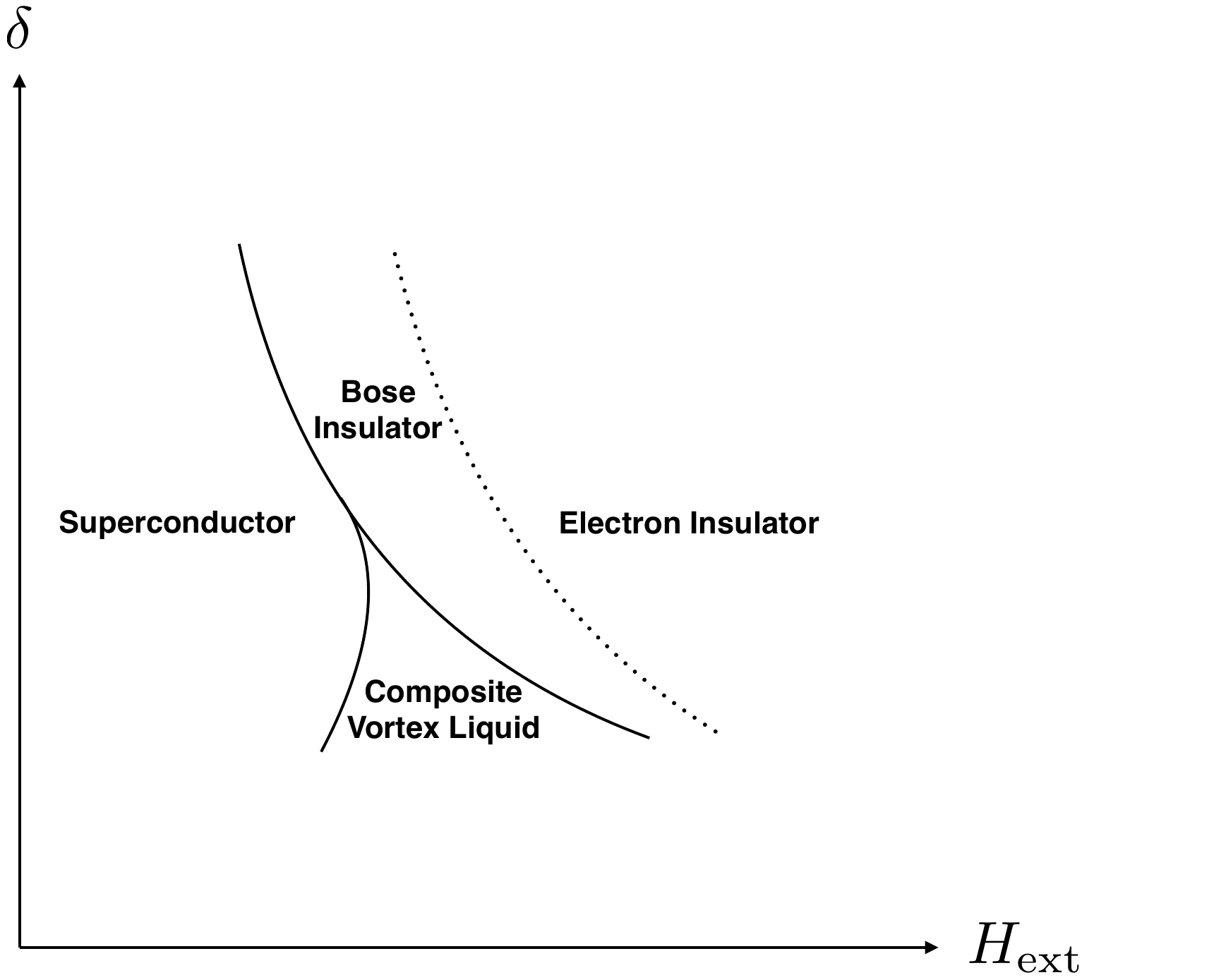}\\
\caption{Schematic $T=0$ phase diagram in the vicinity of the SIT of disordered thin films as a function of external magnetic field $H_{\rm ext}$ and disorder strength $\delta$.
The solid lines denote phase transitions, while the dashed line denotes the boundary (either transition or crossover) between a Bose insulator of Cooper pairs and an electron insulator.
A similar phase diagram obtains for a 2DEG with the relabeling: superconductor $\leftrightarrow$ integer quantum Hall effect, Bose insulator $\leftrightarrow$ Hall insulator \cite{Kivelson1992}, and composite vortex liquid $\leftrightarrow$ composite Fermi liquid \cite{Kalmeyer1992, halperin1993}.
Appendix \ref{phasediagram2} has a second possible phase diagram.}
\label{pd}
\end{figure}

Here, we address - at a phenomenological level - how a metallic phase can emerge starting from the strongly disordered dirty boson limit.  
Our analysis builds on the analogy between direct SITs and quantum Hall plateau transitions, which has been discussed in the literature using a composite boson mapping \cite{Lee1991, Kivelson1992, ZhangIJMP}.  
In contrast, we employ the composite fermion description \cite{Jainbook}, which more naturally accounts for the metallic behavior.
As summarized in Fig. \ref{pd}, we suggest the metallic {\it point} observed at the SIT in dirtier samples is closely related to the metallic {\it phase} in somewhat cleaner systems; in both cases, the metal is analogous to the composite Fermi liquid \cite{Kalmeyer1992, halperin1993} observed in half-filled Landau levels of a 2D electron gas (2DEG)  \cite{Willett1987, Jiang1989}. 
The composite fermions here are field-induced vortices attached to a unit flux quantum of an emergent gauge field.  
The hypothesis that composite fermions determine the low-energy behavior in both systems straightforwardly allows for a number of testable predictions as the two systems complement and inform one another.

\section{Effective Field Theory for the SIT}
We begin with an effective description of a disordered 2D superconductor close to a field-tuned SIT at $T=0$.
The effective Hamiltonian for the delocalized Cooper pairs, $H_{\rm s} = H_1 + H_2$, takes the following form: 
\begin{align}
\label{dirtyH}
H_1 = & \frac{1}{2} \sum_{\bm{r}, \bm{r}'} \Big(n_{\rm s}(\bm{r}) - n_{\rm s} \Big) V_{\bm{r}, \bm{r}'} \Big(n_{\rm s}(\bm{r}') - n_{\rm s} \Big), \cr
H_2 = & - \sum_{i, \bm{r}} J_{i, \bm{r}} \cos\Big( \Delta_{i} \theta(\bm{r}) + e_\ast A_{i}(\bm{r}) \Big).
\end{align}
Here, the unit lattice spacing plays the role of a short-distance cutoff, which is comparable to zero field, zero temperature superconducting correlation length.  
The amplitude of the Cooper-pair field, $\psi \sim e^{i \theta}$, is assumed frozen, while the Cooper-pair density $n_{\rm s}(\bm{r})$ and phase variable $\theta(\bm{r})$ at site $\bm{r} = (x,y)$ satisfy the equal-time commutation relations, $\left[n_{\rm s}(\bm{r}), \theta(\bm{r}') \right] = i \delta_{\bm{r}, \bm{r}'}$.
In Eq. (\ref{dirtyH}), $A_i(\bm{r}) = (A_x(\bm{r}), A_y(\bm{r}))$ is the vector potential for the background magnetic field, $\Delta_i \theta(\bm{r}) \equiv \theta(\bm{r} + \hat{i}) - \bm(\vec{r})$ is the lattice derivative, $V_{\bm{r}, \bm{r}'}$ parameterizes {\it both} the density-density interaction and the coupling to disorder, $J_{i, \bm{r}}$ is the superconducting phase stiffness, which can also vary spatially due to strong disorder, and the Cooper pairs carry electrical charge $e_\ast = 2 e$.

For a clean metal with Galilean invariance at $T=0$, the average Cooper-pair density $\langle n(\bm{r}) \rangle \equiv n_{\rm s} = n_e/2$ where $n_e$ is the electronic carrier density of the metal.  If the same relation were to hold for the disordered films considered here, the field scales required to tune to the SIT would be incompatible with experiment.  
Instead, for a disordered metal at $T=0$, the average Cooper-pair density can be substantially less than $n_e$, since many of the Cooper pairs will be strongly localized, and do not participate in the low-energy effective field theory.  
An estimate of the relevant degrees of freedom can be made via the formula: $n_{\rm s} \sim {n_e \over 3} {\ell^2 \over \hbar D/(k_B T_c)}$ where $\ell$ is the electronic mean-free path, $D$ is the normal-state diffusion constant, and $T_c$ is the zero-field superconducting critical temperature \cite{Abrikosovbook}.
For disordered metals, ${\ell^2 \over \hbar D/(k_B T_c)} \ll 1$ and so the Cooper-pair density may be much reduced from its value in the clean limit.

An important test of the reasoning above is a crude estimation of SIT field scales.
For a 40 $\AA$ thick MoGe film of the type studied by Mason and Kapitulnik, the 3D carrier density $n_e \sim 10^{-26} m^{-3}$, $\ell \sim 6 \times 10^{-10} m$, $D \sim 5 \times 10^{-5} m^2/s$, and $T_c \sim 1 K$ \cite{masonthesis}.  We estimate a 2D Cooper-pair density $n_{\rm s} \sim  2 \times 10^{14} m^{-2}$, which is three orders of magnitude less than the clean estimate of $2 \times 10^{17} m^{-2}$.
The estimated Cooper pair density defines a magnetic field scale $b_{\rm v} = n_{\rm s} \Phi_0 \sim 0.5\ T$
at which the Cooper pairs are at unit filling fraction $\nu = n_{\rm s}/(H_{\rm ext}/\Phi_0)$, where the magnetic flux quantum $\Phi_0 = hc/e_\ast \equiv 2\pi/e_\ast$.
The metallic behavior observed in MoGe films occurs at field scales, which are roughly comparable to  $b_{\rm v}$ \footnote{We have checked that a similar estimate for the Cooper-pair filling fraction where metallic behavior is observed can be made for InO and Ta films via the methods described above, along with an independent estimate using more direct superfluid stiffness measurements in InO films \cite{LiuPanWenKimSamArmitage}, assuming a reduction of the electronic carrier mass from its free value in the range $10^{-1} - 10^{-2}$.}.

Having estimated the average density of mobile Cooper-pairs in a disordered system, we implement the following duality transformation,
\begin{equation}
\label{dualitytrans}
n_{\rm s}(\bm{r}) = {e_\ast \over 2 \pi} b(\bm{r}), \quad \Delta_i \theta(\bm{r}) + e_\ast A_{i}(\bm{r}) = {e_\ast \over 2 \pi} \hat z \times \bm{e}(\bm{r}),
\end{equation}
 {\it only} to these mobile, low energy degrees of freedom.  The duality transformation
relates the density and current operators of the mobile Cooper pairs to the magnetic $b(\bm{r})$ and electric $\bm{e}(\bm{r})$ field of an emergent gauge field that is minimally coupled to the dual vortex degrees of freedom.
The emergent electric field is constrained by Gauss' Law:
\begin{equation}
\Delta \cdot \bm{e}(\bm{r}) = 2 \pi \Big(n_{\rm v}(\bm{r}) - n_{\rm v}\Big),
\end{equation}
where the average vortex density $n_{\rm v} \equiv \langle n_{\rm v}(\bm{r}) \rangle = H_{\rm ext}/\Phi_0$ is sourced by the external magnetic field.
The resulting dual vortex Hamiltonian \cite{fisher1989}, $\tilde{H}_{\rm v} = \tilde{H}_1 + \tilde{H}_2 + \tilde{H}_3$:
\begin{align}
\label{vortexH}
\tilde{H}_1 = & {1 \over 2} \sum_{\bm{r}, \bm{r}'} \Big(n_{\rm v}(\bm{r}) - n_{\rm v} \Big) \tilde{V}_{\bm{r}, \bm{r}'} \Big(n_{\rm v}(\bm{r}') - n_{\rm v} \Big), \cr
\tilde{H}_2 = & - \sum_{i, \bm{r}} \tilde{J}_{i, \bm{r}} \cos\Big(\Delta_i \phi(\bm{r}) - e_\ast a_i(\bm{r})\Big), \cr
\tilde{H}_3 = & \sum_{\bm{r}} \bm{e}(\bm{r}) \cdot \bm{e}(\bm{r}) + \sum_{\bm{r}, \bm{r'}} b(\bm{r}) V_{\bm{r}, \bm{r}'} b(\bm{r}').
\end{align}
In Eq. (\ref{vortexH}), $\phi(\bm{r})$ is the number operator conjugate to $n_{\rm v}(\bm{r})$, $a_i(\bm{r}) = (a_x(\bm{r}), a_y(\bm{r}))$ are the spatial components of the emergent gauge field introduced in Eq. (\ref{dualitytrans}) with average emergent magnetic flux $\langle b \rangle = n_{\rm s} \Phi_0$.
The emergent gauge charge carried by the vortices is equal and opposite to the electromagnetic charge carried by the Cooper pairs.
The vortices are electrically neutral.
$\tilde{V}_{\bm{r}, \bm{r}'}$ and $\tilde{J}_{i, \bm{r}}$ parameterize the density and phase interactions.
Duality implies that the dimensionless electrical resistivity $\bar{\rho}$ is equal to the dimensionless vortex conductivity $(\bar{\rho}^{\rm v})^{-1}$ in units of $h/4e^2$ and {\it vice-versa}.  
As the external field is increased, nucleated vortices, pinned at first by disorder, become mobile at the SIT due to quantum fluctuations, and lead to non-zero resistance.  

\section{Self-Duality and Composite Vortices}
A useful theoretical anchorpoint is the notion of {\it self-duality} which implies $n_{\rm v} = n_{\rm s}$ \footnote{A more precise consequence of self-duality is the relation $\bar{\rho}^{\rm v}_{ij} = \bar{\rho}_{ji}$ which implies $\rho_{xx}^2 + \rho_{xy}^2 = R_Q^2$.}.
Since the supercurrent is defined as  $I = e_\ast \dot n_{\rm s}$, whereas the Josephson relation implies that the voltage, $V= {h \over e_\ast} \dot n_{\rm v}$ in the limit of vanishingly small Hall resistivity, 
a spectacular prediction of self-duality is that the resistance at the SIT must be universal and equal to $R_Q$.
A triumph of the dirty boson theory is the observation of roughly this value for the critical resistance at the SIT in a variety of films \cite{Paalanen1992,Steiner2008,Breznay2015}.

In our phenomenological treatment here, we take the observation of self-duality as an empirical fact and consider its consequences.  
This in turn will enable us to speculate on the nature of the ground state when self-duality is broken by reducing film disorder.    

Since $n_{\rm s} = n_{\rm v}$ at a SIT with self-duality, the transition represents the point where both the Cooper pairs and vortices are on the verge of condensation.
To extend this description to the metallic phase, we must consider the stability of the above picture.
Since Cooper pairs `see' a vortex as a unit of flux and {\it vice-versa}, and because both particles are mutual bosons, it follows that in a mean-field description,
the ground state involves bosons at $\nu=1$.  
However, in the presence of disorder {\it and} interactions, such a mean-field treatment favors the mutually contradictory ground states consisting either of localized Cooper pairs or vortices.

A much more stable mean-field solution is obtained via flux attachment \cite{Jainbook}.
Bosons at $\nu = 1$ map onto a composite fermion metal in zero net background magnetic flux.
Importantly, the Pauli principle is operative in the fermionic description, which substantially stabilizes the mean-field ground state.
Thus, we propose that a composite fermion metal provides an effective description for a self-dual SIT with a finite $T=0$ resistance.
How this picture quantitatively results in the observed experimental behavior requires additional work; nevertheless, it is possible to explore a few of the immediate implications of this hypothesis.


\subsection{Emergent Metallic Phases}
The most important consequence is the possibility that the metallic point broadens into an emergent metallic {\it phase}. 
To be specific, we consider a composite Fermi liquid of composite vortices, which are bosonic vortices bound to a unit flux of the emergent gauge field $a$, whose curl equals the Cooper pair density. 
We propose the following scenario.
At strong disorder, the finite density of composite vortices is driven to a strong disorder fixed point that describes the SIT.
As the disorder is weakened and the effects of a Fermi surface of composite vortices becomes better defined, the fixed point broadens into a metallic phase.

To motivate this hypothesis, we draw upon a well-known analogy \cite{Lee1991} between the SIT and quantum Hall plateau transitions.  
Consider an integer quantum Hall plateau transition from $\nu \rightarrow \nu-1$.  
Upon increasing the external field, the electron chemical potential is lowered, and holes are nucleated in the filled Landau level.  
These holes are the precise analogs of the vortices nucleated by a field in a superconductor.  
The holes are localized until the field is tuned to its critical value at which the plateau transition occurs.  
A consequence of Landau level particle-hole symmetry -- the analog of particle-vortex symmetry -- is that $\sigma_{xy} = - \nu + 1/2$.
In addition, the diagonal conductivity takes a universal value of $e^2/h$ \cite{Shahar1995,Shahar1996}.  
The metallic point that obtains at a plateau transition is closely related to the composite Fermi liquid of half-filled Landau levels.  
In samples with somewhat less disorder, 
there is a metallic phase intervening between the two plateaus \cite{Wong1996}.  
The fact that the metallic point in dirtier 2DEGs broadens into a metallic phase in cleaner 2DEGs reflects the stability of the composite fermion metal relative to either quantum Hall state with Hall conductance of $-\nu$ or $-\nu+1$ (in units of $e^2/h$).  
In the same manner, we suggest here that the metallic phase related to the composite vortex liquid can be a more stable phase than either the superconductor or the insulator.  

\subsection{Composite Vortex Lagrangian}
To test these ideas more quantitatively, we provide an effective Hamiltonian for the composite vortex degrees of freedom, generalizing $\tilde{H}_{\rm v}$.
At unit filling fraction, the composite vortices see zero effective magnetic field, form a Fermi surface \cite{PasquierHaldane, Read1998}, and interact via an emergent gauge field $\tilde{a}$ in addition to the already present field $a$ introduced in $\tilde{H}_{\rm v}$. 
We consider the following ``working" effective Lagrangian, written in the continuum, for the composite vortices $\psi$, ${\cal L}_{\rm cv} = {\cal L}_{\rm 0} + {\cal L}_{\rm gauge} + {\cal L}_{\rm int}$:
\begin{align}
\label{cvlagrangian}
{\cal L}_{{\rm 0}} & = \psi^\dagger \Big(i\partial_t + (\tilde{a}_t - a_t) + {1 \over 2 m_v} (\partial_j - i (\tilde{a}_j - a_j))^2 \Big) \psi, \cr
{\cal L}_{\rm gauge} & = - {1 \over 4 \pi} \epsilon_{\mu \nu \rho} \tilde{a}_\mu \partial_\nu \tilde{a}_\rho + {e_\ast \over 2 \pi} \epsilon_{\mu \nu \rho} A_\mu \partial_\nu a_\rho, \cr
{\cal L}_{\rm int} & = - {1 \over 2} \int d^2 \bm{r}' \Big(\psi^\dagger \psi(\bm{r}) - n_{\rm v} \Big) \tilde{V}_{\bm{r}, \bm{r}'} \Big(\psi^\dagger \psi(\bm{r}') - n_{\rm v} \Big).
\end{align}
We take the composite vortex kinetic term ${\cal L}_{\rm 0}$ to be that of a non-relativistic fermion, applicable to a metallic phase.
The first term in ${\cal L}_{\rm gauge}$ implements the unit flux attachment, while the second term describes the coupling of the Cooper-pair current $J_\mu = {e_\ast \over 2 \pi} \epsilon_{\mu \nu \rho} \partial_\nu a_\rho$ to the external electromagnetic field $A_\mu$. 
The third term ${\cal L}_{\rm int}$ describes the composite vortex density-density interactions inherited from $\tilde{H}_{\rm v}$.
Because the composite fermion $\psi$ only couples to the linear combination $\tilde{A} = \tilde{a} - a$, we may simplify ${\cal L}_{\rm cv}$ and integrate out $a$ to obtain \footnote{We safely ignore the Maxwell terms contained in $\tilde{H}_3$ because they are irrelevant at low-energies compared to the Chern-Simons term in ${\cal L}_{\rm gauge}$.}:
\begin{align}
\label{gauge2}
{\cal L}_{\rm gauge} = - {e_\ast \over 2 \pi} \epsilon_{\mu \nu \rho} \tilde{A}_\mu \partial_\nu A_\rho + {e_\ast^2 \over 4 \pi} \epsilon_{\mu \nu \rho} A_\mu \partial_\nu A_\rho.
\end{align}
The first term in ${\cal L}_{\rm gauge}$ describes the induced coupling between the external gauge field $A$ and the composite vortices through $\tilde{A}$.
The presence of the second term is reminiscent of a composite fermion treatment of the holes -- the analog of the vortices of a superconductor -- in a filled Landau level \cite{BMF} (See also \cite{alicea2005}).

\subsection{Composite Vortex Response}
Consider the consistency of the above scenario with the general phase diagram depicted in Fig. \ref{pd}.
As explained in Appendix \ref{responseappendix}, we can relate the electrical conductivity tensor $\sigma_{ij}$ to the composite vortex conductivity $\sigma_{ij}^{\rm cv}$:
\begin{align}
\sigma_{ij} = {e_\ast^2 \over 2 \pi} \Big(- \epsilon_{jk} + (\sigma^{\rm cv})^{-1}_{jk} \Big).
\end{align}
We concentrate on the DC response as $T \rightarrow 0$.

\subsubsection{Superconductor}
To obtain an electrical superconductor, 
we must take the composite vortices to be insulating, $\sigma_{xx}^{\rm cv}(T \rightarrow 0) = 0$ and $\sigma_{xy}^{\rm cv}(T \rightarrow 0) = 0$.
An Anderson insulator of composite vortices is expected for low composite fermion density when the external magnetic field is small.
The Hall conductivity is sensitive to the rate at which the various components of the composite vortex conductivity tensor vanish: (1) when $\lim_{T \rightarrow 0} {(\sigma_{xx}^{\rm cv})^2 \over \sigma_{xy}^{\rm cv}} \rightarrow 0$, $\sigma_{xy}(T \rightarrow 0)$ diverges, while (2) when $\lim_{T \rightarrow 0} {\sigma_{xy}^{\rm cv} \over (\sigma_{xx}^{\rm cv})^2} \rightarrow 0$, $\sigma_{xy}(T \rightarrow 0)$ is finite.
(Note that we always assume $\lim_{T \rightarrow 0} {\sigma_{xx}^{\rm cv} \over (\sigma_{xy}^{\rm cv})^2} \rightarrow 0$ in order to ensure superconducting behavior.)
The second case may be referred to as a Hall superconductor.
The Hall resistivity, however, vanishes in both cases.

\subsubsection{Insulator}
An electrical insulator
obtains when the composite vortices exhibit the integer quantum Hall effect, $\sigma_{xx}^{\rm cv}(T \rightarrow 0) = 0$ and $\sigma_{xy}^{\rm cv}(T \rightarrow 0) = - 1$.
Analogous to the situation with the superconductor, the precise value of the Hall resistivity depends upon the rates by which the components of the composite vortex conductivity tensor approach their zero temperature values: (1) when $\lim_{T \rightarrow 0} {(\sigma_{xx}^{\rm cv})^2 \over (1+\sigma_{xy}^{\rm cv})} \rightarrow 0$, $\rho_{xy}(T \rightarrow 0)$ diverges, while (2) when $\lim_{T \rightarrow 0} {(1 + \sigma_{xy}^{\rm cv}) \over (\sigma_{xx}^{\rm cv})^2} \rightarrow 0$, $\rho_{xy}(T \rightarrow 0)$ is finite.
(We assume $\lim_{T \rightarrow 0} {\sigma_{xx}^{\rm cv} \over (1 + \sigma_{xy}^{\rm cv})^2} \rightarrow 0$ in order to ensure insulating behavior.)
The first case represents a trivial insulator, while the second case may be called a Hall insulator \cite{Kivelson1992,Zhang1992}.

As parameters in the laboratory are varied, we might expect a crossover from a Hall insulator to a trivial insulator (along with an analogous crossover in the superconducting phase), consistent with the various phases furnished by the composite vortex Lagrangian.

\subsection{Composite Vortices in the Weak Disorder Limit}

We now address the stability of the composite vortex liquid to disorder.
The scaling theory of localization only applies to Fermi liquids coupled to quenched chemical potential disorder.
By contrast, chemical potential disorder in $\tilde{H}_{\rm v}$ translates into {\it random flux} disorder of zero mean and random chemical potential disorder.
To see this, consider the composite vortex Lagrangian in Eq. (\ref{cvlagrangian}) where the coupling of the composite vortex density to random chemical potential disorder is implicitly included in ${\cal L}_{\rm int}$.
The equation of motion for the time-component $\tilde{a}_t$ of the emergent gauge field,
\begin{align}
\label{constraint}
\psi^\dagger \psi(\bm{r}) = {1 \over 2 \pi} \tilde{b}(\bm{r}),
\end{align}
relates the local composite vortex density to the emergent gauge flux $\tilde{b} = \partial_x \tilde{a}_y - \partial_y \tilde{a}_x$, thereby tying fluctuations in the local chemical potential to those of the emergent gauge field \footnote{Recall that the mean-field ansatz for ${\cal L}_{\rm cv}$ is one where $\tilde{b} = b$ with $b = n_{\rm v} \Phi_0$ so the constant offset cancels and the fluctuations in both the chemical potential and flux $\tilde{b}$ are correlated.}.

Happily, the problem of an electron hopping on a lattice in the midst of random chemical potential and random flux of zero mean was studied in \cite{Kalmeyer1992,Kalmeyer1993} with the conclusion that localization is avoided.
Random flux of zero mean is of crucial importance.
Random flux with non-zero average effectively acts as an additional contribution to the random chemical potential and results in localization of non-interacting electrons.
Thus, one of the most important simplifying features of the composite fermion transformation wherein vortices at unit filling fraction map to composite vortices in zero background flux is also one of the most important in guaranteeing stability of the resulting metal to weak disorder.

\section{Experimental Consequences}
We now briefly discuss a few experimental consequences, many of which are readily adapted from the corresponding implications of a composite fermion metal in the 2DEG.

\subsubsection{Heat Capacity}
A simple, but remarkable thermodynamic signature of the metallic phase should, in principle, arise in the heat capacity. 
A linear in temperature heat capacity is expected for a Fermi liquid-like metal; additional corrections due to the interactions of the composite vortices with the Chern-Simons gauge field may be expected to take the form $\delta c_V \sim T^{2/3}$ or $\delta c_V \sim T \log(T)$ depending upon the effective composite vortex density-density interaction.
Distilling such behavior in 2DEGs has thus far proven difficult \cite{Gornik1985, Wang1992}.

\subsubsection{Quantum Oscillations}
Quantum oscillations indicative of the composite vortex Fermi surface with wave vector determined by the applied external magnetic field would provide a striking confirmation of the picture presented in this paper.
The frequency of oscillation will monotonically depend on the deviation of the external field from the value at which the composite vortices experience zero flux.
Precisely this behavior is observed in the 2DEG about half-filling \cite{Jainbook}.

\subsubsection{Tunneling Density of States}
Suppression of the bulk electron tunneling density of states at half-filling in a 2DEG is a direct result of the composite fermion picture \cite{he1993, kim1994}.
However, for the composite vortex metal, a suppression might be expected simply because of the finite superconducting amplitude.
To probe the composite vortex liquid more directly, one should consider instead the tunneling between a superconducting probe and the composite vortex metal, which we expect to be exponentially suppressed.
Within the dirty boson framework, there would be un-suppressed tunneling at the SIT for a superconducting tip.

\subsubsection{Thermopower and Nernst Effect}
Thermopower and the Nernst effect may be utilized to determine whether self-duality is realized in the thin film.
Self-duality, obtained at the SIT or about unit filling fraction of the vortices, requires a vanishing thermopower and Nernst signal \footnote{This conclusion follows from the transformation of the electric field $E_i \mapsto - E_i$ and thermal gradient $(\nabla T)_i \mapsto (\nabla T)_i$ under the particle-vortex transformation. Alternatively, we can see this by mapping the superconductor in zero field to the integer quantum Hall effect where the particle-vortex transformation is traded for the particle-hole transformation. In the quantum Hall problem, the thermopower and Nernst signal are odd under the anti-unitary particle-hole transformation defined in the next footnote.}.
Likewise, particle-hole symmetry within a single Landau level requires the thermopower to vanish 
\footnote{This follows from the transformation of the $E_i \mapsto - E_i$ and $(\nabla T)_i \mapsto (\nabla T)_i$ under the anti-unitary particle-hole transformation generated by ${\cal CT}$ where ${\cal C}$ is charge-conjugation and ${\cal T}$ is time-reversal.
The thermopower remains odd if instead ${\cal C P}$ with ${\cal P}$ the generator spatial reflection along the $y$-axis is identified as the generator of particle-hole transformations.
The Nernst signal, however, is even under ${\cal CP}$.
If a measurement of the thermopower indicated unbroken particle-hole symmetry, a further measurement of the Nernst signal could determine whether the physically-realized particle-hole symmetry were anti-unitary or unitary in the 2DEG.}.
We note that the thermopower is observed to be non-zero for $T > 0$ in 2D hole systems at half-filling in the lowest and first Landau levels \cite{YingBayotSantosShayegan1994, Bayot1995}.
Taken at face value, this indicates a breakdown of particle-hole symmetry.
An obvious guess for the cause of the breakdown is Landau level mixing which is not typically small in the 2DEG; a second possibility is the spontaneous breakdown of particle-hole symmetry in the limit of vanishingly small Landau level mixing.
Thus, if the relation between the 2DEG and thin film film is taken seriously, we may abstract an important lesson: the breakdown of particle-hole symmetry observed in experimentally-realized 2DEGs may imply and provide a mechanism for a similar breakdown of particle-vortex duality in the thin film.

\section{Discussion}
We have suggested that the 2D metallic phase observed in the vicinity of the magnetic field-tuned SIT is a composite vortex liquid, analogous to the composite Fermi liquid found in 2DEGs near half-filling.
This hypothesis entails the statistical transmutation of interacting bosons in a magnetic field into a Fermi sea of composite vortices and provides a natural explanation for the emergent metal with a variety of experimental consequences.



There have been several previous studies formulating a theory of the metallic phase\cite{Das1999,Kapitulnik2001,DalidovichPhillips2001, SpivakOretoKivelson,Galitski2005}. 
The work by Galitski et al. \cite{Galitski2005} whose theoretical formulation most closely resembles ours, identified the metal with a gas of neutral spinons coupled to vortices by an emergent $U(1)$ statistical gauge field \cite{Feigelman1993}.  By contrast, our approach and motivation rely on the observation of self-duality at the field-tuned SIT, the analogy between the SIT and quantum Hall plateau transitions, and the substantial superconducting fluctuations observed within the metal \cite{LiuPanWenKimSamArmitage}.


The superconductor-metal transition has been well studied in thin MoGe and InO films.
It would be of great interest to better understand the nature of the weak insulator in systems where a metal is observed, similar to that in Ta films.
In disordered films where the metallic phase has shrunk to a point, the proximate insulator is a Bose insulator of localized Cooper pairs with non-zero superconducting correlations. 
Upon increasing the magnetic field, the Cooper pairs are broken and an electron insulator results.
By continuity, we expect the Bose insulator to continue to border the emergent metal for as long as the Bose insulator persists as the disorder is decreased as depicted in Fig. \ref{pd}.
A second possible phase diagram is presented in Fig. \ref{pd2} of Appendix \ref{phasediagram2}. 
Here, the superconducting, Bose insulating, electron insulating, and metallic phases meet at a multicritical point that is of higher degree than occurs in Fig. \ref{pd}.
Distinguishing the two possibilities merits further study.

Interestingly, there are indications \cite{Breznay2015} that the Bose insulator observed in InO films is a Hall insulator with vanishing conductivity tensor, diverging linear resistivity, but finite Hall resistivity as $T \rightarrow 0$. 
Hall insulators were, not surprisingly, first predicted \cite{Zhang1992,Kivelson1992} and observed \cite{Goldman1993,Kravchenko1994,Wong1995} in the 2DEG.
Indications of a similar state in the thin film further unites the two systems and gives us further hope for a relation between the metals.  

Several open theoretical questions remain, in light of the observations made here.  Building upon previous work \cite{Chakravarty1986, Chakravarty1988}, further investigation of self-duality in numerical studies of Josephson junction arrays in the presence of strong disorder and magnetic fields is desired.  Second, the emergence of a metallic phase as the disorder strength is reduced would be a clear validation of the ideas presented here.  Furthermore, a deeper understanding of the role of disorder in composite Fermi liquids, taking into account both the random flux disorder due to the Chern-Simons field, as well as non-Fermi liquid effects is needed \cite{disorderongoing}.  Lastly, the application, if any, of recent developments considering particle-hole symmetry -- the analog of particle-vortex duality -- in half-filled Landau levels \cite{Son2015, wangsenthil1, maxashvin2015, Kachru2015}, in the context of the emergent metal near the SIT, is an intriguing possibility.

\acknowledgements We thank P. Armitage, E. Berg, S. Chakravarty, D. Fisher, M. P. A. Fisher, S. Kivelson, R. Laughlin, O. Vafek, A. Vishwanath, J. Yoon, and especially A. Kapitulnik for helpful discussions.  
We thank S. Kachru, R. Laughlin, and C. Nayak for comments on an early draft of this manuscript.
This work was supported in part by  DOE Office of Basic Energy Sciences, contract DE-AC02-76SF00515 (SR) and the John Templeton Foundation (MM).

\appendix

\section{Response Derivation}
\label{responseappendix}

To relate the composite vortex conductivity to the electrical conductivity, we formally integrate out the composite vortices of ${\cal L}_{cv}$, while working in the gauge $\tilde{A}_t = A_t = 0$:
\begin{align}
{\cal L}_{\rm cv} = {1 \over 2} \tilde{A}_j \Pi^{\rm cv}_{jk} \tilde{A}_k + e_\ast {i \omega \over 2 \pi} \epsilon_{jk} \tilde{A}_j A_k - e^2_\ast {i \omega \over 2 \pi} \epsilon_{jk} A_j A_k,
\end{align}
where the composite vortex response $\Pi^{\rm cv}$ and conductivity $\sigma^{\rm cv}$ tensors:
\begin{align}
\Pi^{\rm cv} = {i \omega \over 2 \pi} \begin{pmatrix}
\sigma^{\rm cv}_{xx} & \sigma^{\rm cv}_{xy} \cr - \sigma^{\rm cv}_{xy} & \sigma^{\rm cv}_{xx}
\end{pmatrix}.
\end{align}
Finally, we integrate out $\tilde{A}$ to obtain the effective electronic response Lagrangian,
\begin{align}
{\cal L}_{\rm cv} = {i \omega \over 2} {e^2_\ast \over 2 \pi} A_j \Big(- \epsilon_{jk} + (\sigma^{\rm cv})^{-1}_{jk} \Big) A_k,
\end{align}
from which we may read off the electrical conductivity and resistivity tensors:
\begin{align}
\sigma_{xx} = & {e_\ast^2 \over h} {\sigma_{xx}^{\rm cv} \over (\sigma_{xx}^{\rm cv})^2 + (\sigma_{xy}^{\rm cv})^2}, \cr
\sigma_{xy} = & - {e_\ast^2 \over h} \Big(1 + {\sigma_{xy}^{\rm cv} \over (\sigma_{xx}^{\rm cv})^2 + (\sigma_{xy}^{\rm cv})^2}\Big), \cr
\rho_{xx} = & {h \over e_\ast^2} {\sigma_{xx}^{\rm cv} \over (\sigma_{xx}^{\rm cv})^2 + (1 + \sigma_{xy}^{\rm cv})^2}, \cr
\rho_{xy} = & {h \over e_\ast^2} \Big(1 - {1 + \sigma_{xy}^{\rm cv} \over (\sigma_{xx}^{\rm cv})^2 + (1 + \sigma_{xy}^{\rm cv})^2}\Big).
\end{align}

\section{Alternative Phase Diagram}
\label{phasediagram2}
\begin{figure}[h!]
  \centering
\includegraphics[width=.9\linewidth]{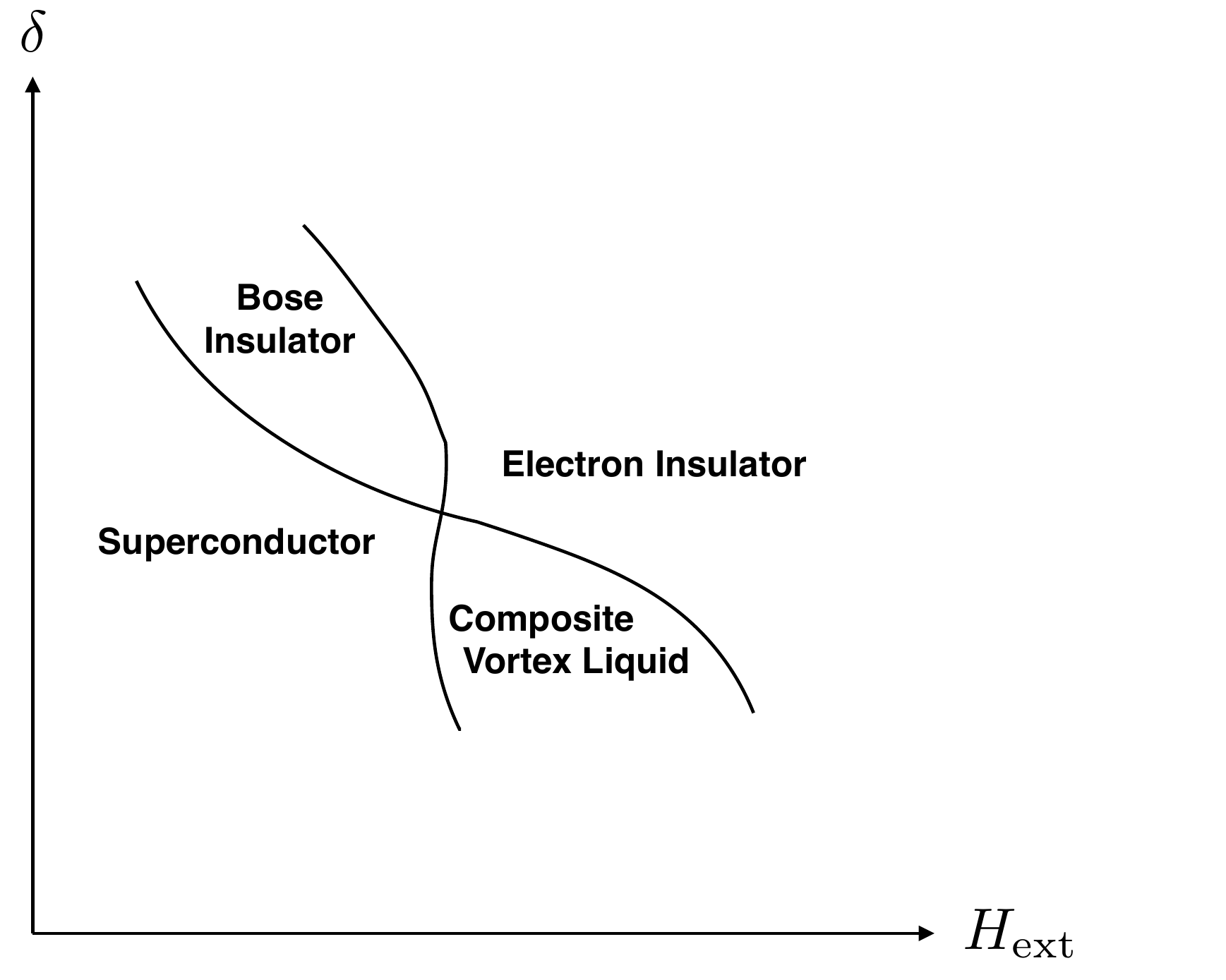}\\
\caption{}
\label{pd2}
\end{figure}
In Fig. \ref{pd}, the superconducting, Bose insulating, and metallic phases meet at a point, while in Fig. \ref{pd2} the same three phases along with the electron insulator meet at a point.
Fig. \ref{pd2} displays a transition between the Bose insulator and electron insulator.
If instead, there happened to be a crossover, the degree of multicriticality at which the various phases meet would be the same in both figures.

\bibliography{sit}

\end{document}